\documentclass[3p]{elsarticle}
\usepackage{amssymb,amsmath}
\usepackage{amsthm}
\usepackage{epsfig}
\usepackage{float}
\usepackage{color}
\usepackage{mathscinet}
\usepackage{pdflscape}
\journal{ } 
\DeclareMathOperator\arctanh{arctanh}
\DeclareMathOperator\cosech{cosech}
\DeclareMathOperator\Li{Li}
\DeclareMathOperator\argcosh{argcosh}

\begin{document}
\begin{frontmatter}

\title{Phase transitions in persistent and run-and-tumble walks}

\author[1]{Karel Proesmans}
\author[2]{Raul Toral}
\author[1]{Christian Van den Broeck}
\address[1]{Hasselt University, B-3590 Diepenbeek, Belgium}
\address[2]{IFISC, Instituto de F{\'\i}sica Interdisciplinar y Sistemas Complejos (UIB-CSIC), Campus Universitat de les Illes Balears, 07122-Palma de Mallorca, Spain}


\begin{abstract}
We calculate the large deviation function of the end-to-end distance and the corresponding extension-versus-force relation for (isotropic) random walks, on and off-lattice, with and without persistence, and in any spatial dimension. For off-lattice random walks with persistence, the large deviation function undergoes a first order phase transition in dimension $d> 5$. In the corresponding force-versus-extension relation, the extension becomes independent of the force beyond a critical value. The transition is anticipated in dimensions $d=4$ and $d=5$, where full extension is reached at a finite value of the applied stretching force. Full analytic details are revealed in the run-and-tumble limit. Finally, on-lattice random walks with persistence display a softening phase in dimension $d=3$ and above, preceding the usual stiffening appearing beyond a critical value of the force. 
\end{abstract}
\end{frontmatter}

%
%
\section{Introduction}
Random walks are everywhere. After playing a prominent role in the early development of statistical mechanics and stochastic processes, they have modeled an incredibly wide array of phenomena in many fields, ranging from sociology and ecology, over economy and finance, to physics and chemistry \cite{pearson,spitzer,montroll1974introduction,fama,berg1993random,weiss,redner2001guide}.
In the basic model, the walker makes a step of a fixed length in a random direction in space. Depending on the problem at hand, additional prescriptions are included. A prominent example is the consideration of persistence, whereby the walker has a preference to make a step in the same direction as its previous one. This takes into account the fact that the motion, while still random, is subject to inertia or memory. A particular limit, called the run-and-tumble walk, has been studied extensively in the context of bacterial dynamics \cite{BB,MS,SMHLF,FHB,CMPT}.
Concomitant with the rich variety of problems, mathematical results have been obtained covering many aspects of random walks. A question of considerable interest is the statistics of the end-to-end distance and the related response properties upon applying a stretching force. In this paper, we report on a surprising first order phase transition in off-lattice random walks with persistence in dimension $d> 5$. In the corresponding force-versus-extension relation, the extension becomes independent of the force beyond a critical value. The transition is anticipated in dimensions $d=4$ and $d=5$, where full extension is reached at a finite value of an applied stretching force. Full analytic details are revealed in the run-and-tumble limit.

\section{Interpretation of the large deviation function}
The asymptotic statistics of the end-to-end distance along a given coordinate, say $X$, is typically described by a so-called large deviation function (LDF) \cite{Keller}. More precisely, we focus on the statistics of the scaled distance $x=X/(Nb)$, with $N$ the number of steps, $b$ the step size. In the limit $N\rightarrow \infty$, the corresponding probability $P(x)$ assumes the asymptotic form \cite{Hollander,Donsker}: 
\begin{equation}\label{LDF}
P(x)\sim \exp\{-N \mathcal{I}(x)\},
\end{equation}
with the LDF $\mathcal{I}(x)$ a non-negative and typically convex function. It is equal to zero in the ``overwhelmingly most probable" value $\bar{x}$ which, for an unbiased walk corresponds to $\bar{x}=0$.
For example, the LDF for a symmetric random walk in $1$ dimension reads
 $\mathcal{I}(x)=[{(1+x)}\ln(1+x)+{(1-x)}\ln(1-x)]/2$.
The limiting values $\mathcal{I}(\pm 1)=\ln2$ reproduce the probability $1/2^{N}$ for always stepping in the same direction. 
 
 A revealing interpretation of the large deviation function can be given if we assume that the realizations of the random walk correspond to equally probably (iso-energetic) states of a system (e.g., a polymer) in equilibrium. Invoking the microcanonical definition (Boltzmann-Einstein formula) for the entropy $\mathcal{S}(X)=k_B \ln W(X)$, where $W(X)$ is the number of realizations giving rise to $X$ and $k_B$ is Boltzmann's constant, we find that $P(x)\sim W(X) \sim \exp \{[\mathcal{S}({X})-\mathcal{S}(\bar{X})]/k_B\}$ ($\bar{X}=N\bar{x}$). Comparing with Eq.(\ref{LDF}), one can make the following asymptotic identification:
$\mathcal{S}(\bar{X})-\mathcal{S}(X)\sim N k_B \mathcal{I}(x)$.
 Within this setting, it is clear how to explore the region of exponentially unlikely realizations, namely by introducing a (constant) biasing force $F$ along the stretching direction. The appropriate thermodynamic potential, namely the Gibbs free energy $\mathcal{G}(F)$ as a function of the external force (and the temperature $T$), is given by $\mathcal{G}(F)=\min_X\{-T\mathcal{S}(X)-FX\}=N \min_x\{k_B T \mathcal{I}(x)-b F x\}$ \cite{tp}. We will denote the corresponding minimum, i.e., the most likely extension for a given force by $\bar{x}_F$ (with $\bar{x}_0=\bar{x}$):
 \begin{equation}\label{minF}
 \mathcal{I}'(\bar{x}_F)= \beta b F\;\;\mbox{or} \;\; \bar{x}_F=\mathcal{L}(\beta b F),
 \end{equation}
where $\beta=1/k_B T$. Here, we have introduced the inverse function of $\mathcal{I}'$, which we call the (generalized) Langevin or response function $\mathcal{L}$.
Note also the appearance of the scaled force $
f\equiv\beta b F$,
which is the work delivered per step divided by the thermal energy $k_B T$. 
In many cases, the Langevin function has a simpler analytic form than the LDF, while it is in principle readily accessible by probing the extension versus force relation. For example, for the above mentioned $1d$ lattice random walk, one has that $\mathcal{L}(f)=\tanh f$. 
The Langevin function can also be obtained\cite{touchettelarge2009} from the scaled cumulant generating function $\mathcal{C}(f)$:
\begin{eqnarray}\label{SCG}
\mathcal{C}(f)= \lim_{N\rightarrow \infty} \frac{1}{N}\ln \langle e^{f X}\rangle=\max_x\{f x-\mathcal{I}(x)\}.
\end{eqnarray}
 Indeed, $\mathcal{C}(f)$ being the Legendre-Fenchel transform of $\mathcal{I}(x)$, the derivatives $\mathcal{C}'$ and $\mathcal{I}'$ are each other inverse function, hence $\mathcal{C}'(f)=\mathcal{L}(f)$.
 Note finally that $\mathcal{C}$ is, in view of its definition in Eq.~(\ref{SCG}), always a convex function, while $\mathcal{I}$ needs not be so. Equilibrium statistical physics however tells us that apparent non-concavity of the entropy, corresponding to non-convexity of the LDF, signals the presence of a phase transition. 
 In this case, one needs to perform a Maxwell's construction, which at the level of the LDF is equivalent to the consideration of its convex envelope. The linear segment of the envelope corresponds to the coexistence of the two phases each represented by one of the endpoints, and the slope gives the value of the relevant intensive parameter (in this case, the force $f$) that stays constant during the coexistence transition. Our crucial discovery is that such a first order phase transition occurs for persistent off-lattice random walks in dimensions $d>5$, with the newly appearing phase corresponding to the macroscopic manifestation of a ``single microstate", namely that of persistent flight segments.

In the next sections, we evaluate the LDF $\mathcal{I}(x)$ of the end-to-end distance and the corresponding Langevin function $\mathcal{L}(f)$ for isotropic random walks, on and off-lattice, with and without persistence, and in any spatial dimension. These results are obtained from the scaled cumulant generating function Eq.(\ref{SCG}). A separate section is devoted to the run-and-tumble model understood as a suitable limit of the off-lattice walks.

\section{Random walks on a lattice}\label{AppA}
\subsection{The non-persistence case}\label{sub:LNP}
Let us consider a (probably biased) random walk in a regular $d$-dimensional lattice. Each time step, a random direction $i=1,2,\dots,2d$ is chosen and the walker moves one lattice step $b=1$ in that direction. We focus of the probability distribution of the final coordinate $X$ after $N$ steps. When, with probability $1/d$, a movement along the X axis has been chosen in a given time step, then the X coordinate increases by one with probability $p$ or decreases by one with probability $1-p$ (we really do not care about what happens in the other directions). We ask for the probability $P(n)$ that the X coordinate is $n$ after $N$ time steps. If we denote by $N_x$ the number of moves in the X-direction, then we have:
\begin{eqnarray}
P(n)&=&\sum_{N_x=0,\,N_x=n (\text{mod }2)}^NP(N_x)P(n|N_x),\\
P(N_x)&=&
{N \choose N_x}\left(\frac{1}{d}\right)^{N_x}\left(1-\frac{1}{d}\right)^{N-N_x},\\
P(n|N_x)&=&{N_x\choose{\frac{N_x+n}{2}}}p^{\frac{N_x+n}{2}}(1-p)^{\frac{N-N_x}{2}}.
\end{eqnarray}
Using Stirling's approximation $\ln N!\approx N \ln N-N$, introducing $u=N_x/N$, $x=n/N$ and replacing the sum over $N_x$ by an integral over $u$ with $du=2/N$, we find
\begin{eqnarray}
P(n)&=&\int_0^1\frac{Ndu}{2}e^{-N\phi_d(u,x)},\\ \nonumber
\phi_d(u,x)&=&\ln d+(1-u)\ln\left(\frac{1-u}{d-1}\right)+\frac{u+x}{2}\ln\left(\frac{u+x}{2p}\right)+\frac{u-x}{2}\ln\left(\frac{u-x}{2(1-p)}\right).
\end{eqnarray}
In the large $N$ limit, we use the saddle-point technique to compute the integral to obtain $P(n)\sim e^{-N{\mathcal I}_d(x)}$ with the large deviation function (LDF) given by ${\mathcal I}_d(x)=\phi_d(u^*(x),x)$ where $u^*(x)$ is the maximum of $\phi_d(u,x)$ for given $x$:
\begin{equation}\label{usx}
u^*(x)=\frac{(d-1)\sqrt{(d-1)^2x^2+4p(1-p)(1-x^2)}-4p(1-p)}{d(d-2)+(1-2p)^2},
\end{equation}
which in the symmetric case, $p=1/2$, simplifies to:
\begin{equation}\label{usx1}
u^*(x)=\begin{cases}1, & d=1,\\ \dfrac{1+x^2}{2},&d=2,\\ \dfrac{(d-1)\sqrt{1+d(d-2)x^2}-1}{d(d-2)},&d>2.\end{cases}
\end{equation}
The explicit expression of the large deviation function for $d=1$ is
\begin{eqnarray}
{\mathcal I}_1(x)&=&\frac{1}{2}\left[(1-x)\ln\left(\frac{1-x}{2(1-p)}\right)+(1+x)\ln\left(\frac{1+x}{2p}\right)\right].
\end{eqnarray}
For $d=2,3$ we just quote the result for the symmetric case, $p=1/2$:
\begin{eqnarray}
{\mathcal I}_2(x)&=&(1-x)\ln(1-x)+(1+x)\ln(1+x),\\
{\mathcal I}_3(x)&=&\ln(3(1-x^2))+x\ln\left(\frac{2x+\sqrt{1+3x^2}}{1-x}\right)-\frac12\ln(5+3x^2+4\sqrt{1+3x^2}\,).\nonumber
\end{eqnarray}
Note that ${\mathcal I}_2(x)=2{\mathcal I}_1(x)$ if $p=1/2$.

The response function $x={\mathcal L}_d(f)$ resulting from ${\mathcal I}_d'(x)=f$, or
\begin{equation}
f={\mathcal I}_d'(x)=\left.\frac{\partial \phi_d(u,x)}{\partial u}\right|_{u=u^*}\frac{\partial u^*}{\partial x}+\frac{\partial \phi_d(u^*,x)}{\partial x}=\frac12\ln\left(\frac{u^*+x}{u^*-x}\right),
\end{equation}
where we have used $\left.\frac{\partial \phi_d(u,x)}{\partial u}\right|_{u=u^*}=0$, by the own definition of $u^*$. Replacing $u^*(x)$ from Eq.(\ref{usx}) we find, after a lengthy but straightforward algebra, a particularly simple result:
\begin{equation}
{\mathcal L}_d(f)=\frac{\sinh (f-f_0)}{\cosh(f-f_0)+(d-1)\cosh(f_0)},\quad f_0=\arctanh(1-2p).
\end{equation}
As ${\mathcal C}_d'={\mathcal L}_d, \,{\mathcal C}(0)=0$, the cumulant generating function is 
\begin{equation}\label{eq:Cf0}
{\mathcal C}_d(f)=\ln\left(\dfrac{\cosh({f-f_0} )+(d-1)\cosh(f_0)}{d\cosh(f_0)}\right).
\end{equation}
It is noticeable that the response curve for arbitrary asymmetry $p$ can be related to that of a symmetric random walk in a different dimension ${\mathcal L}_d(f,p)={\mathcal L}_{d'}(f-f_0,p=1/2)$ with $d'=1+(d-1)\cosh(f_0)=1+(d-1)/({2\sqrt{p(1-p)}})$.

We note the physically expected properties: $\lim_{f\rightarrow \pm\infty}\mathcal{L}_d(f)=\pm 1$ (full extension for infinite force) and, in the symmetric case $p=1/2$ (or $f_0=0$), 
$\mathcal{L}_d(f)=-\mathcal{L}_d(-f)$ (opposite force gives opposite extension). The latter symmetry condition implies $\mathcal{L}_d(0)=0$ (no extension in absence of a force) and $\mathcal{L}_d''(0)=0$, i.e., $\mathcal{L}_d(f)$ has an inflection point at $f=0$. As a result, in the absence of another pair of inflection points, the Langevin function will describe the stiffening (increasing "spring constant" corresponding to a decreasing value of $\mathcal{L}_d'$) upon increasing force. In fact, we obtain from the series expansion
\begin{equation}
{\mathcal L}_d(f)=\frac{f}{d}+\frac{d-3}{6d^2}f^3+\frac{30-15d+d^2}{120d^3}f^5+O(f^7),
\end{equation}
showing that ${\mathcal L}_d(f)$ has, in this case, additional inflection points, besides the one at $f=0$, if $d>d_c=3$.

\subsection{Random walks on a lattice with persistence}\label{sec:rw_persistence}

We next turn to isotropic on-lattice random walks in a regular $d$-dimensional lattice but now we include {\bfseries persistence}\cite{PRW0,PRW1,PRW01,PRW02,PRW2,PRW3,PRW4,PRW5}. Persistence is introduced as a probability $\mathcal{P}>0$ that the walker makes a step in the same direction as the previous one; otherwise, with probability $1-\mathcal{P}$, a random orientation is chosen (which possibly includes the previous orientation). More specifically: at step $k$ there is a probability ${\mathcal P}>0$ that the direction that was taken at step $k-1$, say direction $i\in[1,2d]$, is kept or, otherwise, a new direction is chosen from {\bfseries all } possible directions (including again $i$). 

Let $x_k$ be the increase in position in the $X$ coordinate at step $k$. We are interested in quantifying the probabilities of $3$ possible outcomes $x_k=-1,0,1$ meaning, respectively, a move of $-1$ in the X-coordinate, a move in a direction different from the X axis, and a move of $+1$ in the X-coordinate. The transition matrix of the Markov chain is:
\begin{eqnarray} \label{wpall}
W&=&\begin{pmatrix}
W(-1\to -1)&W(0\to -1)&W(1\to -1)\\
W(-1\to 0)&W(0\to 0)&W(1\to 0)\\
W(-1\to +1)&W(0\to +1)&W(1\to +1)
\end{pmatrix}\\ \nonumber &=&
\begin{pmatrix}
{\mathcal P}+\dfrac{1-{\mathcal P}}{2d} & \dfrac{1-{\mathcal P}}{2d}&\dfrac{1-{\mathcal P}}{2d} \\ (1-{\mathcal P})\left(1-\frac{1}{d}\right) & \quad {\mathcal P}+(1-{\mathcal P})\left(1-\frac{1}{d}\right)\quad & (1-{\mathcal P})\left(1-\frac{1}{d}\right)\\ \dfrac{1-{\mathcal P}}{2d}&\dfrac{1-{\mathcal P}}{2d} &{\mathcal P}+\dfrac{1-{\mathcal P}}{2d}\end{pmatrix},
\end{eqnarray}
with $W(b\to a)=P(x_k=b|x_{k-1}=a)$.

The probability distribution after $N$ steps $\vec P_N=\begin{pmatrix}P(x_N=-1)\\P(x_N=0)\\P(x_N=+1)\end{pmatrix}$, follows from the recursion equation:
\begin{equation}\label{pevol}
\vec P_{k+1}=W\vec P_k \Rightarrow \vec P_N=W^N\vec P_0.
\end{equation}
The average vaue of the cumulant generating function ${\mathcal C}_d(f)$
\begin{equation}\label{eq:Cfrommu}
{\mathcal C}_d(f)=\lim_{N\to \infty}\frac{1}{N}\ln\left\langle e^{f\sum_{k=1}^Nx_k}\right\rangle,
\end{equation}
can be computed as
\begin{eqnarray}
\left\langle e^{f\sum_{k=1}^Nx_k}\right\rangle&=&\sum_{x_0=\pm 1}\sum_{x_1=\pm1}\dots\sum_{x_N=\pm1}e^{fx_N}W(x_N,x_{N-1})e^{fx_{N-1}}W(x_{N-1},x_{N-2})\cdots e^{fx_1}W(x_1,x_0)P_0(x_0)\nonumber\\
&=&\sum_{x_0=\pm 1}\sum_{x_1=\pm1}\dots\sum_{x_N=\pm1}\tilde W(x_N,x_{N-1})\tilde W(x_{N-1},x_{N-2})\cdots \tilde W(x_1,x_0)P_0(x_0).
\end{eqnarray}
Where\footnote{A more symmetric choice: $\tilde W(x_k,x_{k-1})=e^{\frac{f}{2} (x_k+x_{k-1})}W(x_k,x_{k-1})$ gives the same results.} $\tilde W(x_k,x_{k-1})=e^{fx_k}W(x_k,x_{k-1})$, or
\begin{equation}\label{W-persis1}
\tilde W=
\begin{pmatrix}
\left({\mathcal P}+\dfrac{1-{\mathcal P}}{2d}\right) e^{-f}& \dfrac{1-{\mathcal P}}{2d}e^{-f} & \dfrac{1-{\mathcal P}}{2d}e^{-f} \\ (1-{\mathcal P})\left(1-\frac{1}{d}\right) & {\mathcal P}+(1-{\mathcal P})\left(1-\frac{1}{d}\right)& (1-{\mathcal P})\left(1-\frac{1}{d}\right)\\ \dfrac{1-{\mathcal P}}{2d}e^f& \dfrac{1-{\mathcal P}}{2d}e^f& \left({\mathcal P}+\dfrac{1-{\mathcal P}}{2d}\right)e^f
\end{pmatrix}.
\end{equation} 

According to Eq.(\ref{eq:Cfrommu}), if $\mu_d(f)$ is the largest eigenvalue of $\tilde W$, then ${\mathcal C}_d(f)=\ln (\mu_d(f))$. We can in principle find analytically the largest eigenvalue of $\tilde W$ solving a third degree equation, using Cardano's formula, although the resulting expression is long, and not very helpful. Let us mention, for the sake of completeness, the case of non-persistence case ${\mathcal P}=0$ with bias $p$ in the +X direction. The previous matrix now reads:
\begin{equation}\label{W-nopersis-bias}
\tilde W({\cal P}=0)=
\begin{pmatrix}
\dfrac{1-p}{d} e^{-f}& \dfrac{1-p}{d} e^{-f}& \dfrac{1-p}{d} e^{-f}\\ 1-\dfrac{1}{d} & 1-\dfrac{1}{d} & 1-\dfrac{1}{d} \\ \dfrac{p}{d} e^f& \dfrac{p}{d} e^f & \dfrac{p}{d}e^f,
\end{pmatrix}.
\end{equation} 
whose eigenvalues are $0$ (double) and $\left({\cosh({f-f_0} )+(d-1)\cosh(f_0)}\right)/\left({d\cosh(f_0)}\right)$, obtaining in a much simpler way the result Eq.(\ref{eq:Cf0}) of subsection \ref{sub:LNP}.

Another important simplification occurs in $d=1$. In this case the $0$ state does not exist and the problem can be simplified by limiting ourselves to the $2\times 2$ matrix:
\begin{equation}
\begin{pmatrix}\frac{1+{\mathcal P}}{2}e^{-f}& \frac{1-{\mathcal P}}{2}e^{-f} \\ \frac{1-{\mathcal P}}{2} e^f&\frac{1+{\mathcal P}}{2}e^f
\end{pmatrix}.
\end{equation} 
After finding the largest eigenvalue of this matrix, we obtain:
\begin{equation}
{\mathcal C}_1(f)=\ln\left[\frac{1+{\mathcal P}}{2}\left(\cosh(f)+\sqrt{\cosh^2(f)-\dfrac{4{\mathcal P}}{(1+{\mathcal P})^2}}\right)\right],
\end{equation}
and a response function
\begin{equation}
{\mathcal L}_1(f)={\mathcal C}_1'(f)=\frac{\sinh(f)}{\sqrt{\cosh^2(f)-\dfrac{4{\mathcal P}}{(1+{\mathcal P})^2}}}.
\end{equation}
To find the large deviation function we need to perform the Legendre-Fenchel transform: first invert ${\mathcal C}_1'(f)=x$ to find $f(x)$ as
\begin{equation}
\cosh\left[f(x)\right]=\sqrt{\dfrac{1-cx^2}{1-x^2}},\quad c\equiv\dfrac{4{\mathcal P}}{(1+{\mathcal P})^2}.
\end{equation}
Out of the two possible solutions for $f(x)$ we must take $\text{sign}(f(x))=\text{sign}(x)$. The LDF is
\begin{equation}
{\mathcal I}_1(x)=xf(x)-{\mathcal C}_1(f(x)),
\end{equation}
or the explicit expression
\begin{equation}
{\mathcal I}_1(x)=|x|\argcosh\left(\sqrt{\frac{1-cx^2}{1-x^2}}\right)-\ln\left[\frac{1+{\mathcal P}}{2}\left(\sqrt{\frac{1-c}{1-x^2}}+\sqrt{\frac{1-cx^2}{1-x^2}}\right)\right]
\end{equation}
The limiting value is ${\mathcal I}_1(x=\pm 1)=\ln\left(\dfrac{2}{1+{\mathcal P}}\right)$.

We derive now an argument that provides an alternative expression for the eigenvalues of $\tilde W$. The idea is to start from the recursion relation that gives $P_N(X,i)$, the probability that, after $N$ steps, the location of the random walk is $X$ and has been reached from a step in the direction $i=1,\dots,M=2d$. In the case of persistence the recursion relation is:
\begin{eqnarray}
P_N(X,i)&=&{\mathcal P} P_{N-1}(X-x_i,i)+\frac{1-{\mathcal P}}{M}\sum_{j=1}^M P_{N-1}(X-x_i,j),
\end{eqnarray}
where $x_i$ is the variation in the $X$ coordinate of the position of the walker occurs when the direction $i$ has been taken. For the generating function $F_N(f,i)$ we obtain:
\begin{eqnarray}
F_N(f,i)\equiv \int dX\,e^{fX}P_N(X,i)={\mathcal P} e^{fx_i}F_{N-1}(f,i)+\frac{1-{\mathcal P}}{M}e^{fx_i}\sum_{j=1}^M F_{N-1}(f,j).
\end{eqnarray}
The trial function $F_N(f,i)=\mu_d^N\Psi(f,i)$ leads to the eigenvalue problem:
\begin{eqnarray}
\mu_d \Psi(f,i)={\mathcal P} e^{fx_i}\Psi(f,i)+\frac{1-{\mathcal P}}{M}e^{fx_i}\sum_{j=1}^M \Psi(f,j).
\end{eqnarray}
We now introduce in this expression the (arbitrary) normalization condition $\frac{1}{M}\sum_{i=1}^M\Psi(f,i)=1$ to obtain
\begin{eqnarray}
\Psi(f,i)=\frac{1-{\mathcal P}}{\mu_d e^{-fx_i}-{\mathcal P}}
\end{eqnarray}
and using again the normalization condition we obtain a closed equation for the eigenvalues $\mu_d$
\begin{eqnarray}\label{eigenvalues-mu}
\frac{1-{\mathcal P}}{M}\sum_{i=1}^M\frac{1}{\mu_d e^{-fx_i}-{\mathcal P} }=1.
\end{eqnarray}
In the case of a $d$ dimensional regular lattice, one direction contributes $+1$ to the variable $X$, another direction contributes $-1$ and the remaining $2d-2$ do not contribute, such that the above equation reads:
\begin{eqnarray}\label{eigenvalues-mu-lattice}
\frac{1-{\mathcal P}}{2d}\left[\frac{1}{\mu_d e^{-f}-{\mathcal P} }+\frac{1}{\mu_d e^f-{\mathcal P} }+\frac{2d-2}{\mu_d -{\mathcal P} }\right]=1.
\end{eqnarray}
It is possible to check, either numerically or comparing the resulting equations, that the eigenvalues $\mu_d$ obtained from this equation coincide exactly with the eigenvalues of matrix $\tilde W$ as given by Eq.(\ref{W-persis1}).

In summary, for a regular $d$-dimensional lattice with persistence, ${\mathcal P}>0$, once $\mu_d(f)$ has been obtained using either the eigenvalues of matrix Eq.(\ref{W-persis1}) or the solutions of Eq.(\ref{eigenvalues-mu-lattice}), the generalized Langevin function can be obtained as ${\mathcal L}_d(f)={\mathcal C}_d'(f)={\mu_d'(f)}/{\mu_d(f)}$. In practice, we have used the symbolic program Mathematica\cite{Mathematica} to obtain the largest eigenvalue $\mu_d(f)$ and its derivative. This allows us to plot the desired ${\mathcal L}_d(f)$ depicted in Fig.1.

\begin{figure}[h]
\centering
\includegraphics[width=0.75\linewidth]{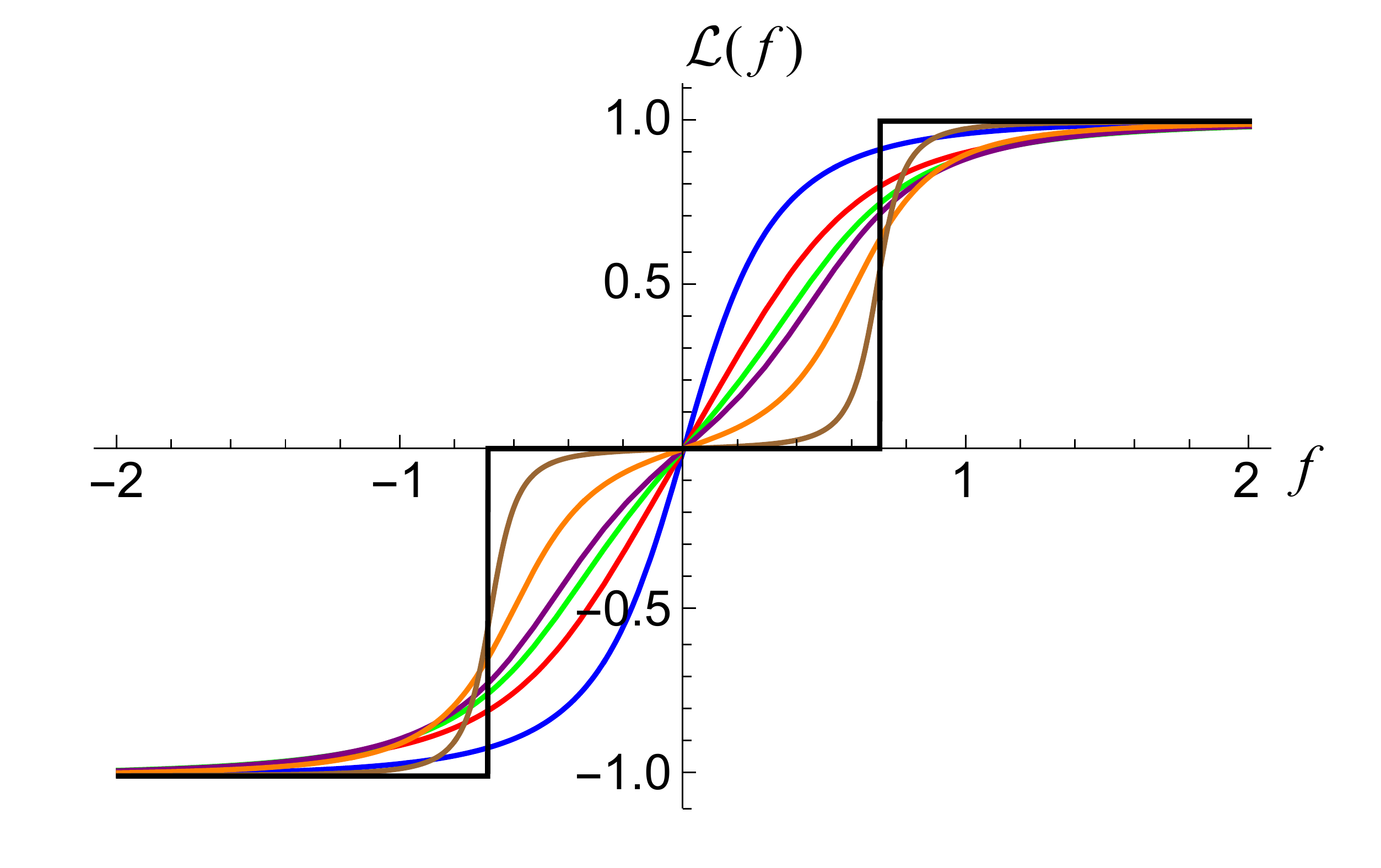}
\caption{Langevin function $\mathcal{L}(f)$ for on-lattice random walk with persistence ${\mathcal P}=0.5$, for $d=1,2,3,4,10,100,\infty$ (in order of decreasing slope at the origin). Note the additional inflection points for $d>2$.}
\label{fig:LP}
\end{figure}

The existence of inflection points other than $f=0$ of the function ${\mathcal L}_d(f)$ can be derived from the expansion of the Langevin function. Expanding the largest eigenvalue in powers of $f$,
\begin{equation}
\mu_d(f)=1+\dfrac{1+{\mathcal P}}{1-{\mathcal P}}f^2+\frac{(1+{\mathcal P})(d(1+10{\mathcal P}+{\mathcal P}^2)-6{\mathcal P}(3+{\mathcal P}))}{	24d^2(1-{\mathcal P})^3}f^4+O(f^6),
\end{equation}
we obtain 
\begin{equation}
{\mathcal L}_d(f)=\frac{1+{\mathcal P}}{d(1-{\mathcal P})}f+\frac{(1+{\mathcal P})(d(1+10{\mathcal P}+{\mathcal P}^2)-3(1+6{\mathcal P}+{\mathcal P}^2)}{6d^2(1-{\mathcal P})^3}f^3+O(f^5).
\end{equation}
Therefore, the equation ${\mathcal L}_d''(f)=0$ has real solutions $\pm f_m\ne 0$ for $d>d_\text{c}=\left({3(1+6{\mathcal P}+{\mathcal P}^2)}\right)/\left({1+10{\mathcal P}+{\mathcal P}^2}\right)$. As ${\mathcal P}\in(0,1)$, it is $d_\text{c}\in(2,3)$ and, limiting ourselves to integer values of $d$, the transition dimension is $d_\text{c}=2$, i.e. for $d\le 2$, ${\mathcal L}_d''$ is single peaked, while it is double peaked for $d>2$.
As a result, a regime of softening around $f=0$, followed by the ``usual" stiffening regime upon further increase of the force beyond a critical amplitude (corresponding to locations of the inflection points away from zero) will be observed in on-lattice random walks for $d>3$, and from $d=3$ in the case of persistence.

The Langevin function ${\mathcal L}_d(f)$ is an analytical function of its argument $f$ for all finite values of the dimension $d$. However, in the mathematical limit $d\to\infty$ a singularity does appear. As one can derive from Eq.\eqref{eigenvalues-mu-lattice} we obtain the values
\begin{equation}\label{eq:eigen_d=inf}
\mu_\infty(f)=\begin{cases}
{\mathcal P} e^{-f}, & f< \ln {\mathcal P},\\ 
1, & \ln {\mathcal P} < f< -\ln{\mathcal P},\\ 
{\mathcal P} e^{-f}, & f< \ln {\mathcal P},
\end{cases}
\end{equation}
and
\begin{equation}\label{eq:eigen_d=inf2}
{\mathcal L}_\infty(f)= {\mathcal C}_\infty'(f)=\dfrac{\mu_\infty'(f)}{\mu_\infty(f)}=\begin{cases}
-1, & f < \ln {\mathcal P},\\ 
0, & \ln {\mathcal P} < f < -\ln{\mathcal P},\\ 
1, & f < \ln {\mathcal P},
\end{cases}
\end{equation}
with singularities at $f=\pm\ln{\mathcal P}$. 
From here, the equation ${\mathcal L}_\infty(f)=x$ has the solution
\begin{equation}
\begin{cases}
f=\ln{\mathcal P}, &\text{if }x\in(-1,0),\\
f=-\ln{\mathcal P}, &\text{if }x\in(0,1).
\end{cases}
\end{equation}
The large deviation function ${\mathcal I}_\infty(x)=xf(x)-{\mathcal C}_\infty(f(x))$, or
\begin{equation}
{\mathcal I}_\infty(x)=\begin{cases}
x \ln{\mathcal P}, &\text{if }x\in(-1,0),\\
-x\ln{\mathcal P}, &\text{if }x\in(0,1).
\end{cases}
\end{equation}
A curve with a singularity at $x=0$. Interpreting ${\mathcal L}_\infty(f)$ as the response function giving the elongation to an external forcing, it turns out that the elongation is $0$ up to $f=-\ln({\mathcal P})$ and then it jumps to $x=1$, a discontinuous transition, see Fig.\ref{fig:LP}. In the case of an off-lattice random walk with persistence, considered in the next sections, the same type of discontinuities will be observed at finite values of the dimension $d$.

\section{Random walks off-lattice}\label{AppB}

We consider now that the random walker moves in ${\mathbb R}^d$ and executes a movement of length $b=1$ in a random direction. Let $\theta$ be the angle of the trajectory with the X-direction. We are interested in the distribution of $x=\frac{1}{N}\sum_{k=1}^Nx_k=\frac{1}{N}\sum_{k=1}^N\cos(\theta_k)$ which verifies a large-deviation relation of the form $P(x)\sim e^{-N{\mathcal I}_d(x)}$.

Persistence can be included as in the lattice case by including the probability ${\mathcal P}$ that the direction taken at step $k$ coincides with the one taken at step $k-1$. If, with probability $1-{\mathcal P}$, that same direction is not taken then a new direction is taken again randomly. As the angles expand a continuous range of real values, except in $d=1$, the probability that the new direction coincides with the previous one is zero. Note also that, if $d>1$ the only way of reaching $x=1$ is by repeatedly iterating an initial move in the $X$ direction, an event that happens with a probability $P(x=\pm 1)\sim{\mathcal P}^N=e^{N\ln {\mathcal P}}$, which leads to $I_d(x=\pm 1)=-\ln {\mathcal P}$ for $d>1$.

We do not need to develop the theory anew. If we allow the direction angle to change continuously in dimension $d$, we can obtain the Legendre-Fenchel transform of the large deviation function ${\mathcal C}_d(f)=\log\mu_d(f)$ by extending Eq.\eqref{eigenvalues-mu} to the case of $M\to\infty$ taking $x_i=\cos(\theta)$ and the corresponding angular distribution 
\begin{equation}
g_d(\theta)=\dfrac{\Gamma\left(\frac{d}{2}\right)}{\sqrt{\pi}\Gamma\left(\frac{d-1}{2}\right)}\sin^{d-2}(\theta), \hspace{20pt}\int_0^\pi d\theta\,g_d(\theta)=1.
\end{equation}
Hence, $\mu_d(f)$ is found as the solution of the equation,
\begin{eqnarray}\label{Fd:eq}
\int_0^\pi d\theta \frac{g_d(\theta)}{\mu_d e^{-f\cos(\theta)}-{\mathcal P}} =\frac{1}{1-{\mathcal P}}.
\end{eqnarray}
We analyze this equation first in the non-persistence case ${\mathcal P}=0$ and then in the general case of a non-null persistence.

\subsection{Random walks off-lattice without persistence}

Setting ${\mathcal P}=0$ in Eq.(\ref{Fd:eq}), we obtain
\begin{eqnarray}
\mu_d(f)&=&\int_0^\pi d\theta\,g_d(\theta)e^{f\cos(\theta)}=\left(\frac{2}{f}\right)^{\frac{d}{2}-1}\Gamma\left(\frac{d}{2}\right)I_{\frac{d}{2}-1}(f),
\end{eqnarray}
being $I_\nu(f)$ the hyperbolic Bessel function of order $\nu$. The response function is ${\mathcal L}_d(f)={\mathcal C}_d'(f)={\mu_d'(f)}/{\mu_d(f)}$ or
\begin{equation}
{\mathcal L}_d(f)=\frac{I_{\frac{d}{2}}(f)}{I_{\frac{d}{2}-1}(f)}.
\end{equation}
This includes the well-known results $\mathcal{L}_1(f)=\tanh f$ ($d=1$) , and $\mathcal{L}_3(f)=\coth f-1/f$ ($d=3$) \cite{kuhn}. 
The Taylor expansion 
\begin{equation}
{\mathcal L}_d(f)=\frac{f}{d}-\frac{f^3}{d^2 (2 + d)} + \frac{2 f^5}{d^3 (2+d)(4+d)}+O(f^7)
\end{equation}
indicates that the third derivative of ${\mathcal L}_d(f)$ at $f=0$ is always negative and there are no further inflection points other than $f=0$.

Except for $d=1$ where the result coincides with the lattice case, the corresponding LDF's for $d> 1$ do not have a simple analytic expression. We note, however, that for large $d$, one can use the asymptotic expression of the Bessel function\footnote{See, for example, {\tt https://dlmf.nist.gov/10.41\#iv}}:
\begin{equation}
\ln I_\nu(z)\sim \sqrt{z^2+\nu^2}+\nu\ln\left(\frac{z}{\nu+\sqrt{\nu^2+z^2}}\right),\, z\to \infty.
\end{equation}
to obtain the asymptotic result
\begin{equation}\label{ixlim}
{\mathcal I}_d(x)\sim -\dfrac{d}{2}\ln(1-x^2).
\end{equation}

\subsection{General case of non-null persistence}\label{subsec:non-null}

Let us introduce $\xi_d\equiv\mu_d/{\mathcal P}$. From Eq.(\ref{Fd:eq}), $\xi_d(f)$ is found by solving 
\begin{eqnarray}\label{Fd:eq1}
F_d(\xi_d,f)&=&\frac{{\mathcal P}}{1-{\mathcal P}},\\
F_d(\xi,f)&\equiv&\int_0^\pi d\theta \frac{g_d(\theta)}{\xi e^{-f\cos(\theta)}-1}.\label{Fd:eq2}
\end{eqnarray}

The problem with this equation is that, as we will see, the solution $\xi_d(f)$ might not exist for some values of $f$ and $d$. This does not make sense as $f$ is an arbitrary parameter taking any possible value in $\mathbb{R}$. To see what goes wrong, we first need to analyze the integral in some detail. 

-Note that $F_d(\xi,f)=F_d(\xi,-f)$ is symmetric with respect to $f$. Therefore, we restrict ourselves to the interval $f\in[0,\infty)$. 

-For the integral to be convergent the conditions $\xi>e^f$ and $\xi<e^{-f}$ must be satisfied, otherwise the denominator becomes $0$ at some value of $\theta$ and the integral does not exist. If $\xi<e^{-f}$ the integrand is always negative and $F_d(\xi,f)$ can never be equal to ${{\mathcal P}}/({1-{\mathcal P}})\in[0,\infty)$. Therefore, we need $\xi>e^f$ (recall we are considering only $f\ge 0$). 

-$F_d(\xi,f)$ is a monotonously decreasing function of $\xi\ge 0$ taking its maximum value at $\xi=e^f$ (the minimum allowed value for $\xi$). If this maximum value $F_d(e^f,f)$ is finite then, for a given ${\mathcal P}$ there exists a maximum possible value $f_\text{c}$ for which the solution $\xi_d(f)$ exists. If this is the case, the maximum value $f_\text{c}(d,{\mathcal P})$, which depends on dimension $d$ and persistence probability ${\mathcal P}$, is found by solving 
\begin{equation}
F_d(\xi=e^{f_\text{c}},f_\text{c})=\dfrac{{\mathcal P}}{1-{\mathcal P}}.
\end{equation}
Only if $F_d(\xi=e^f,f)=\infty,\,\forall f$, will Eqs.~(\ref{Fd:eq1},\ref{Fd:eq2}) have a solution for all values of $f$. As we will see, this is the case if $d\le 3$.

Besides $d=1$, where the problem is identical to the lattice case, a particularly simple case is $d=3$ where the integral can be expressed in terms of elementary functions:
\begin{eqnarray}
F_3(\xi,f)&=&\dfrac{1}{2f}\ln\left(\frac{\xi- e^{- f}}{\xi- e^{ f}}\right)
\end{eqnarray}
leading to 
\begin{eqnarray}
\xi_3(f)&=&\dfrac{\sinh\left[f/({1-{\mathcal P}})\right]}{\sinh\left[{\mathcal P} f/(1-{\mathcal P})\right]},\\ 
{\mathcal C}_3(f)&=&\ln\left[{\mathcal P}\, \xi_3(f)\right],\\
{\mathcal L}_3(f)&=&{\mathcal C}_3'(f)=\dfrac{\coth\left[f/(1-{\mathcal P})\right]-{\mathcal P}\coth\left[{\mathcal P}f/(1-{\mathcal P}]\right)}{1-{\mathcal P}}.\end{eqnarray}
Note that $\xi_3(f)$ exists for any value of $f$ and that ${\mathcal L}_3(f\to\pm\infty)=\pm1$. For $d\ge 5$ odd it is possible to find some analytical expressions for ${\mathcal L}_d(f)$, see \ref{analytical}, but the calculation of $\xi_d(f)$ for a given value of the persistence probability $\mathcal P$ has to be performed by a numerical solution of Eqs.(\ref{Fd:eq1},\ref{Fd:eq2}). For $d$ even, the whole determination of $\xi_d(f)$, ${\mathcal C}_d(f)$ and ${\mathcal L}_d(f)$ has to be done numerically.

Recall that ${\mathcal L}_d(f)\in[-1,1]$, and it is important to determine the value of this response function at $f=f_\text{c}$. If ${\mathcal L}_d(f_\text{c})<1$, then the equation ${\mathcal L}_d(f)=x\in[-1,1]$ can not always be solved. This failure has a formal analogy in the calculation of the partition function of an ideal quantum boson gas\cite{Pathria-3rd}, that we now outline briefly. For a Bose-Einstein ideal gas the thermodynamic potential $J(z,T,V)$ as a function of the fugacity $z$, the temperature $T$ and the volume $V$ is obtained in the grand-canonical ensemble as $J=-k_BT\sum_{\ell}\ln(1-ze^{-\varepsilon_\ell/k_BT})$, where $\varepsilon_\ell$ is the energy of the single-particle energy quantum level $\ell$. Using the non-relativistic density of states in three dimensions, $g(\varepsilon)=\frac{2\pi V(2m)^{3/2}}{h^3}\varepsilon^{1/2}$, the potential $J$ is computed as an integral $J=-k_BT\int_0^{\infty}d\varepsilon g(\varepsilon)\ln(1-ze^{-\varepsilon/k_BT})=k_BTV\lambda_T^{3/2}\Li_{5/2}(z)$, with $\lambda_T=\frac{h}{\sqrt{2\pi m k_BT}}$ and the polylogarithm appears as the integral $\Li_s(\xi)=\frac{1}{\Gamma(s)}\int_0^\infty dx {x^{s-1}}/({\xi^{-1} e^x-1})$. The problem arises when the equation of state that follows from this thermodynamic potential, $N=V\lambda_T^{-3}\Li_{3/2}(z)$, can not be right as it predicts that the maximum possible value of the density is $\frac{N}{V}=\lambda_T^{-3}\Li_{3/2}(1)=2.612\lambda_T^{-3}$, and it does not make sense to have an upper limit for the density of particles.

The solution is well known: in passing from a sum to an integral using the density of states $g(\varepsilon)$ the contribution of the ground state $\varepsilon=0$ is completely lost as $g(0)=0$, but at low enough temperatures the bosons condensate in the ground state (Bose-Einstein transition) and its contribution to the sum can not be neglected. This is solved by including explicitly the contribution of the ground state by making the replacement of a sum by an integral in the following way:
\begin{equation}
\sum_{\ell}\ln(1-ze^{-\varepsilon_\ell/k_BT})=\int_0^{\infty}d\varepsilon g(\varepsilon)\ln(1-ze^{-\varepsilon/k_BT})+\ln(1-z).
\end{equation}

The failure of our calculation can be solved in a similar way: We realize that the integral, Eq.~(\ref{Fd:eq}) comes from a sum Eq.(\ref{eigenvalues-mu}) in the limit of $M\to\infty$ terms in the sum, but that in the limit process of replacing the sum by an integral we have lost the contribution of $\theta=0$ as the weighting factor $\sin(\theta)^{d-2}$ completely disregards its contribution for $d>2$. As in Bose-Einstein theory, the solution is to include the $\theta=0$ term explicitly, replacing Eqs.(\ref{Fd:eq1}-\ref{Fd:eq2}) by 
\begin{eqnarray}\label{Fd:eq3}
F_d^M(\xi,f)\equiv\frac{1}{M}\frac{1}{\xi e^{-f}-1}+\int_0^\pi d\theta \frac{g_d(\theta)}{\xi e^{-f\cos(\theta)}-1} =\frac{{\mathcal P}}{1-{\mathcal P}}.
\end{eqnarray}
The equation $F_d^M(\xi_d,f)=\dfrac{{\mathcal P}}{1-{\mathcal P}}$ now has a finite solution for all $f$ as long as $M$ is finite. When $M\to\infty$ the solution for $f>f_\text{c}$ tends to $\xi_d=e^f$. This implies that ${\mathcal C}_d(f)=\ln({\mathcal P}\xi_d)=f +\ln {\mathcal P}$ and ${\mathcal L}_d(f)=1$ for $f>f_\text{c}$. The inclusion of the explicit $\theta=0$ summand is needed as long as $f_\text{c}$ is finite, for dimension $d>3$. As a consequence, the function ${\mathcal L}_d(f)$ develops a singularity at $f=f_\text{c}$. The type of singularity depends on the dimension. For $3<d\le 5$ the function ${\mathcal L}_d(f)$ is continuous at $f=f_\text{c}$, but its derivative is not, whereas for $d >5$ the function itself has a finite discontinuity. This singularity propagates into the cumulant generating function ${\mathcal C}_d(f)$ and also into the large deviation function ${\mathcal I}_d(x)$ that develops a singularity at $x=x_\text{c}={\mathcal L}_d(f_\text{c})$, that is at $x_\text{c}<1$ for $d>5$.

\begin{figure}[h]
\centering
\includegraphics[width=0.6\linewidth]{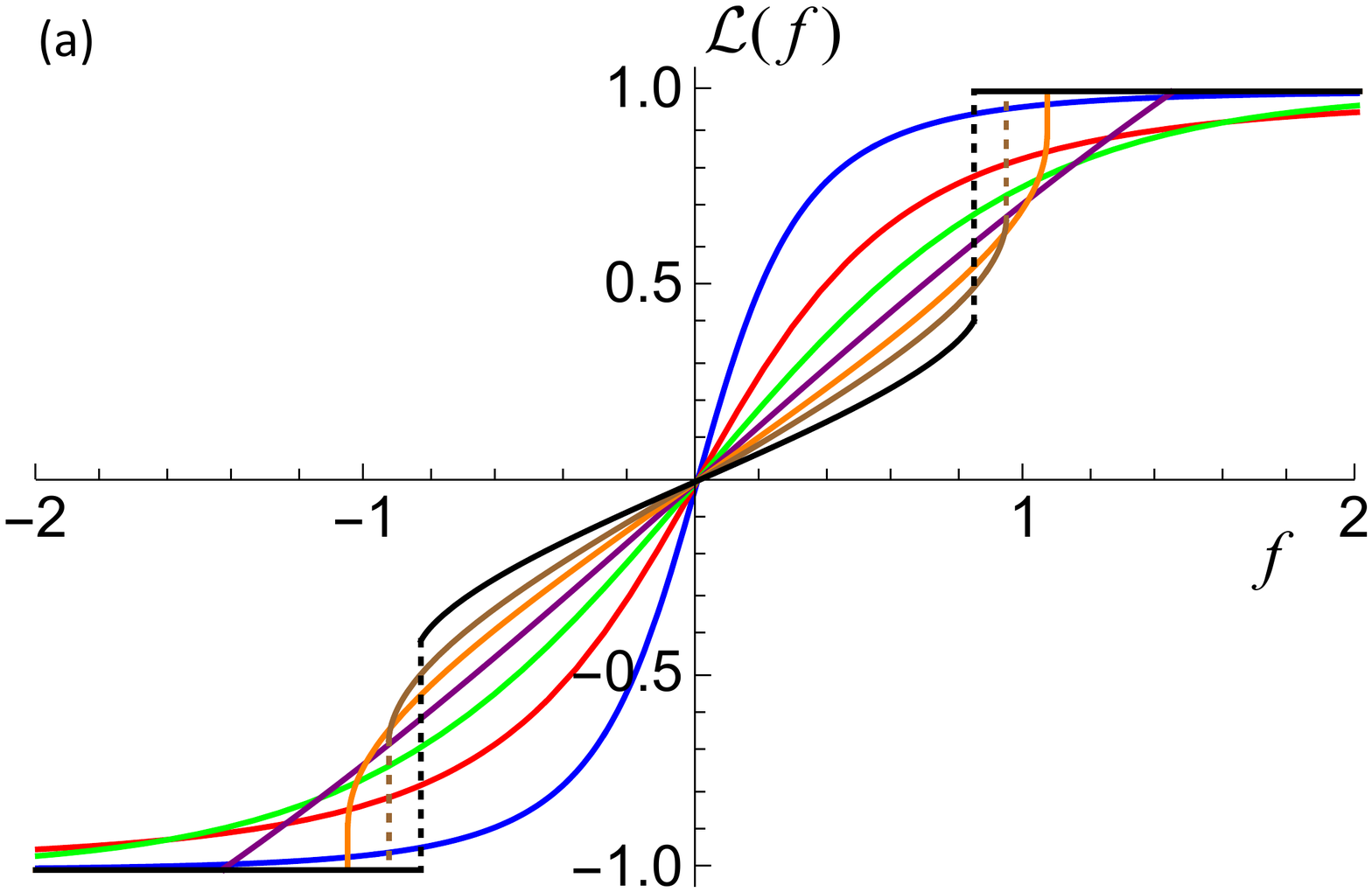}
\includegraphics[width=0.6\linewidth]{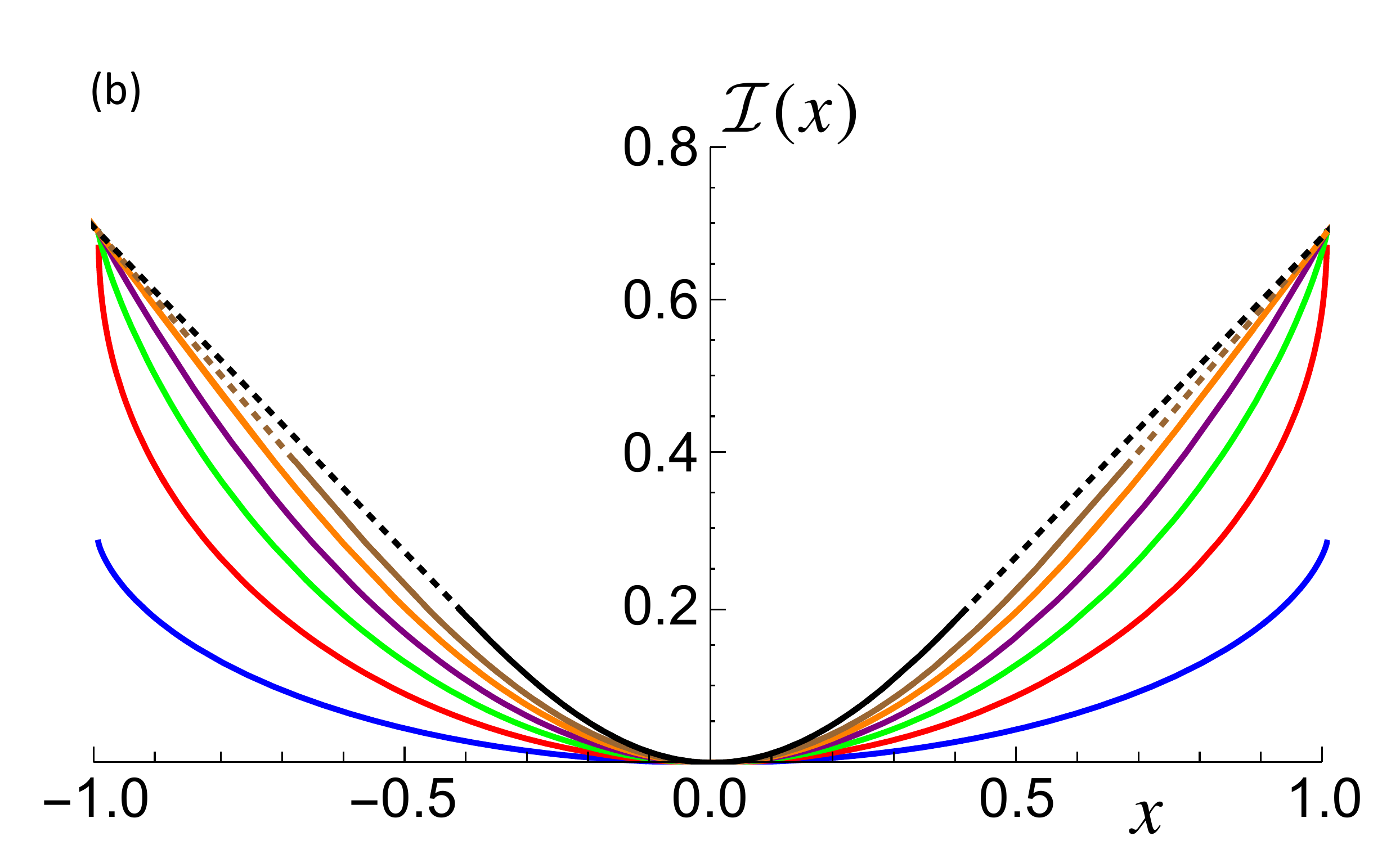}
\caption{a) Langevin function $\mathcal{L}(f)$ for $d=1,2,3,4,5,6,8$ (from decreasing slope at the origin) and b) large deviation function $\mathcal{I}(x)$ (same color code) for an off-lattice random walk with persistence ${\mathcal P}=0.5$. Full extension is reached at a finite critical value $f_\text{c}$ of the force for $d=4$ and $d=5$, while a discontinuous transition, as indicated by dotted vertical lines in panel (a), takes place for $d>5$.}
\label{fig:OP}
\end{figure}

Using the explicit expressions above we can prove that $f_\text{c}$ decreases with increasing dimension. The value of $\xi_\text{max}=\xi(f_\text{c})=e^{f_\text{c}}$, also decreases with dimension. For instance, if ${\mathcal P}=0.5$ we have the numerical values:
\begin{table*}[h]
\begin{center}
\begin{tabular}{|c|c|c|c|c|}
\hline
$d$ & $f_\text{c}$ &${\mathcal L}_d(f_\text{c})$\\
\hline
$\le 3$ & $\infty$ &1\\
4& 1.44266&1\\
5 & 1.06073&1\\
6 & 0.936596&0.67822\\
7 & 0.87513&0.512275\\
8 & 0.838444&0.411353\\
9 & 0.814067&0.343574\\
\hline
\end{tabular}
\end{center}
\end{table*}\\
From this table, we note that the value of $f_\text{c}$ decreases steadily with dimension (diverging at $d=3$), but the maximum value of ${\mathcal L}_d$ reaches $1$ up to $d=5$ and then it takes values less than $1$. Generalized Langevin functions ${\mathcal L}_d(f)$ for different values of spatial dimension $d$ and a persistence value ${\mathcal P}=0.5$, have been plotted in Fig.\ref{fig:OP}(a).

Given the aforementioned properties of ${\mathcal C}_d(f)$, it turs out that its Legendre-Fenchel transform, the LDF ${\mathcal I}_d(x)$ develops an inflection point at $x=x_\text{c}={\mathcal L}_d(f_\text{c})$ and becomes non-convex for $x\in[x_\text{c},1]$. As mentioned earlier, the right interpretation is to introduce the Maxwell construction, and replace the function ${\mathcal I}_d(x)$ in that interval by the straight line connecting the points $(x_\text{c},{\mathcal I}_d(x_\text{c}))$ and $(1,-\ln{\mathcal P})$:
\begin{equation}
{\mathcal I}_d(x)=\dfrac{{\mathcal I}_d(x_\text{c})+\ln{\mathcal P}}{x_\text{c}-1}(x-1)-\ln{\mathcal P},\quad \text{ for }~ d\ge 5 \text{ and } x\in(x_\text{c},1),
\end{equation}
or, using ${\mathcal I}_d(x_\text{c})=x_\text{c}f_\text{c}-{\mathcal C}(f_\text{c})=f_\text{c}(x_\text{c}-1)-\ln{\mathcal P}$,
\begin{equation}
{\mathcal I}_d(x)=f_\text{c}(x-1)-\ln{\mathcal P},\quad \text{ for }~ d\ge5 \text{ and } x\in(x_\text{c},1).
\end{equation}
This Maxwell construction has been used when plotting the LDF of Fig.\ref{fig:OP}(b) for $d=6,8$.

\section{Run-and-tumble model}\label{AppC}

We have introduced persistence in a discrete-time random process by considering that there is a finite probability ${\mathcal P}$ that the previous direction $\theta$ is kept at each time step. It is possible to introduce persistence in a continuous-time version of the random walk, the so-called run-and-tumble model, which might be more natural. We still consider that initially a random direction $\theta\in[0,\pi]$ in ${\mathbb R}^d$ is chosen from the distribution $g_d(\theta)$. That direction is then followed by the walker during a time $\tau$ chosen from an exponential distribution $e^{-\tau/\tau_0}$, time after which a new random direction is chosen  independently of the previous one, being $\tau_0$ a characteristic time. Before the new change of direction, the variable $X$ has varied in $v_0\tau\cos\theta$, with $v_0=\ell_0/\tau_0$ the speed of the particle, and the time $t$ increased by $\tau$. After a large time, the probability distribution to find an X-coordinate $X$ at time $t$ will be shown to follow a large deviation function $P(X,t)\sim e^{-t{\mathcal I}_d(X/t)}$. Note that the corresponding LDF $\mathcal{I}_d(x)$ now has limiting values at $x=\pm1$, corresponding to an uninterrupted straight flight with probability $\exp(-N)$, i.e., $\mathcal{I}_d(\pm1)=1$ (except in $d=1$ where $\mathcal{I}_1(\pm1)=1/2$ due to the finite state-space), instead of the random walk value ${\cal I}(\pm1)=\ln 2$. There are several ways in which this large deviation function ${\mathcal I}_d(x)$ can be obtained. The simplest one is by a limit process that we describe in the next subsection. Other ways of getting the LDF are explained in \ref{alternative}. Without lack of generality we set $\ell_0=\tau_0=v_0=1$.

\subsection{Limit process}

%
Probably the simplest way is to relate the continuous-time and the discrete-time versions of the model is by a limit process. To this end, we consider a discrete-time random walk in which every time step $i=1,\dots, N$, of duration $dt$, the current direction is kept with probability ${\mathcal P}=1-dt$, and a new direction $\theta_i$ is chosen otherwise. During the time $dt$ the walker moves a distance $dx=dt$. In the limit $dt\to 0$, this is equivalent to saying that a direction is kept during a finite time $\tau$ drawn from an exponential distribution $e^{-\tau}$, or during a distance $\ell$ is drawn from a distribution $e^{-\ell}$. In the discrete case we know that $\hat X\equiv \sum_{i=1}^N\cos(\theta_i)$ follows a large deviation form $P(\hat X,N)\sim e^{-N{\mathcal I}_d(\hat X/N)}$. In the continuous-time version the distance followed by the random walker is $X =\sum_{i=1}^Ndt \cos(\theta_i)=dt \hat X$. As the time of the continuous random walk is $t=Ndt$ it follows that the probability of $X$ can be written as $e^{-N{\mathcal I}_d(\hat X/N)}=e^{-N{\mathcal I}_d(X/Ndt)}=e^{-t{\mathcal I}_d(X/t)}$ with ${\mathcal I}_d(x)={\mathcal I}_d(x)/dt$, or, more properly ${\mathcal I}_d(x)=\lim_{dt\to 0}{\mathcal I}_d(x)/dt$ (remember that $dt$ appears also in the numerator within the persistence probability ${\mathcal P}=1-dt$). Concerning the relation between the Legendre-Fenchel transforms of both functions, a simple calculation leads to ${\mathcal C}_d(f)=\lim_{dt\to 0}{\mathcal C}_d(fdt)/dt$.

In Eq.(\ref{Fd:eq}), after using the definition $\mu_d(f)=e^{{\mathcal C}_d(f)}$, we replace $f\to f dt$, ${\mathcal P}\to1-dt$, ${\mathcal C}_d(fdt)\to dt \,{\mathcal C}_d(f)$ and take the limit $dt\to 0$, obtaining 
\begin{equation}\label{eq:Cf1}
1=\int d\theta\,\dfrac{g_d(\theta)}{1+{\mathcal C}_d(f)-f\cos(\theta)}.
\end{equation}


\subsection{The large deviation function}
We use Eq.(\ref{eq:Cf1}) to compute the cumulant generating function ${\mathcal C}_d(f)$ and to obtain its Legendre-Fenchel transform ${\mathcal I}_d(x)$. The integral can be expressed in terms of the Gauss hypergeometric function $F(\alpha_1,\alpha_2;\alpha_3;x)$ also written in the literature\cite{AbramowitzStegun} as ${_2F_1}(\alpha_1,\alpha_2;\alpha_3;x)$:
\begin{eqnarray}\label{eq:Fdz}
1=\int_0^\pi d\theta\,\dfrac{g_d(\theta)}{a-f\cos(\theta)}&=&\frac{1}{a}F\left(\frac12,1;\frac{d}{2};\left(\frac{f}{a}\right)^2\right),\quad |f|<|a|\end{eqnarray}
where $a=1+{\mathcal C}_d(f)$.
As the hypergeometric function $F\left(\frac12,1;\frac{d}{2};z\right)$ can be expressed for integer $d$ in terms of elementary functions, it is possible in some cases to give explicit expressions for ${\mathcal C}_d(f)$, its derivative ${\mathcal L}_d(f)$ and the LDF ${\mathcal I}_d(x)$, as detailed in Table \ref{table} for $d=1,2,4,6$. In those cases that an explicit analytical expression is not available, one can always use parametric expressions. For instance, in Eq.(\ref{eq:Fdz}) introduce $z=f/a$ and recall that $\mathcal{C}_d=a-1$. The cumulant generating function can be expressed as:
\begin{eqnarray}
\mathcal{C}_d(z)&=&-1+F\left(\frac12,1;\frac{d}{2};z^2\right),\\
f(z)&=&z F\left(\frac12,1;\frac{d}{2};z^2\right).
\end{eqnarray}
The parametric form of the response function follows from:
\begin{equation}\label{eq:param6}
{\cal L}_d(z)=\dfrac{\frac{\partial \mathcal{C}_d(z)}{\partial z}}{\frac{\partial f(z)}{\partial z}}=\dfrac{2z F\left(\frac32,2;1+\frac{d}{2};z^2\right)}{dF\left(\frac12,1;\frac{d}{2};z^2\right)+2z^2 F\left(\frac32,2;1+\frac{d}{2};z^2\right)}
\end{equation}
where we have used the derivative of the hypergeometric function $F'(\alpha_1,\alpha_2;\alpha_3,x)=\frac{\alpha_1\alpha_2}{\alpha_3}F(1+\alpha_1,1+\alpha_2;1+\alpha_3;x)$.
For the LDF we use $x={\cal L}_d(f)$ and $\mathcal{I}_d(x)=x f-\mathcal{C}_d(f)$, to obtain its parametric form:
\begin{eqnarray}
x(z)&=&{\cal L}_d(z),\\
\mathcal{I}_d(z)&=&x(z)f(z)-\mathcal{C}_d(z).
\end{eqnarray}
For example, for $d=3$ we obtain:
\begin{eqnarray}
x(z)&=&\frac{1}{z}+\left(1-\frac{1}{z^2}\right)\arctanh(z),\label{eq:param7}\\
{\mathcal I}_3(z)&=&1+\left(1-\frac{1}{z^2}\right)(\arctanh(z))^2.\label{eq:param8}
\end{eqnarray}
We also mention the  parametric form of the LDF for $d=5$;
\begin{eqnarray}
x(z)&=&\frac{3(2-z)z+(6-6z+z^2)\ln(1-z)}{z\left(2z+(2-z)\ln(1-z)\right)},\label{eq:param2}\\
 \mathcal{I}_5(z)&=&1+\frac{3\left(\left(2-z\right)z+2(1-z)\ln(1-z)\right)^2}{2z^3(2z+(2-z)\ln(1-z))}.\label{eq:param1}
 \end{eqnarray}
From this, one can calculate the parametric form of the cumulant generating function and the response function:
\begin{eqnarray}
 f(z)&=&\dfrac{\frac{\partial \mathcal{I}_5}{\partial z}}{\frac{\partial x}{\partial z}}=\frac{3(2-z)z+6(1-z)\ln(1-z)}{2z^2},\label{eq:param3}\\
\mathcal{C}_5(z)&=&x(z)f(z)- \mathcal{I}_5(z)=\frac{z (12 + (-12 + z) z) + 6 (-2 + z) (-1 + z) \ln(1 - z)}{2 z^3},\label{eq:param4}\\
 \mathcal{L}_5(f)&=&\dfrac{\frac{\partial \mathcal{C}_5(z)}{\partial z}}{\frac{\partial f}{\partial z}}=\frac{3(z-2)z-(6-6z+z^2)\ln(1-z)}{z(-2z+(z-2)\ln(1-z))}.\label{eq:param5}
\end{eqnarray}
These results are summarized in Table.\ref{table}

\pagestyle{empty}
\begin{landscape}
\begin{table}
\begin{tabular}{|c|c|c|c|c|}
\hline
$\,d\,$&$F(1/2,1,d/2,z^2)$&${\mathcal C}_d(f)$&${\mathcal L}_d(f)$&${\mathcal I}_d(x)$\\
\hline
& & & &\\
$1$&$\dfrac{1}{1-z^2}$&$\dfrac{\sqrt{1+4f^2}-1}{2}$&$\dfrac{2f}{\sqrt{1+4f^2}}$&$\dfrac{1-\sqrt{1-x^2}}{2}$\\
& & & &\\
\hline
& & & &\\
$2$&$\dfrac{1}{u}$&$\sqrt{1+f^2}-1$&$\dfrac{f}{\sqrt{1+f^2}}$&$1-\sqrt{1-x^2}$\\
& & & &\\
\hline
& & & &\\
$3$&$\dfrac{\arctanh(z)}{z}$&$f\coth(f)-1$&$\coth(f)-f\cosech^2(f)$& Eqs.(\ref{eq:param7},\ref{eq:param8})\\
& & & &\\
\hline
& & & &\\
$4$&$\dfrac{2}{1+u}$&$\dfrac{f^2}{4}$&$\dfrac{f}{2}$&$x^2$\\
& & & &\\
\hline
& & & &\\
$5$&$\dfrac{3(z-(1-z^2)\arctanh(z))}{2z^3}$ & Eqs.(\ref{eq:param3},\ref{eq:param4})& Eqs.(\ref{eq:param3},\ref{eq:param5})& Eqs.(\ref{eq:param2},\ref{eq:param1})\\
& & & &\\
\hline
& & & &\\
$6$& $\dfrac{4(1+2u)}{3(1+u)^2}$ &$\frac{2}{3}\left(\cos\left(\frac{w}{3}\right)+2\sin\left(\frac{w}{6}\right)-1\right)$ & $\dfrac{4}{3}\cos\left(\frac{\pi+\arccos\left(\frac{3f}{4}\right)}{3}\right)$& $\dfrac{3}{16}x^2(8-3x^2)$\\
& & & &\\
\hline
\end{tabular}
\caption{\label{table}The response function ${\mathcal L}_d(f)$ and the LDF function ${\mathcal I}_d(x)$ for different values of the dimension $d$. We use the notation $u\equiv\sqrt{1-z^2}$ and $w\equiv \pi+4\arccos(3f/4)$ whenever necessary.}
\end{table}
\end{landscape}
\pagestyle{plain}

\begin{figure}[h]
\centering
\includegraphics[width=0.6\linewidth]{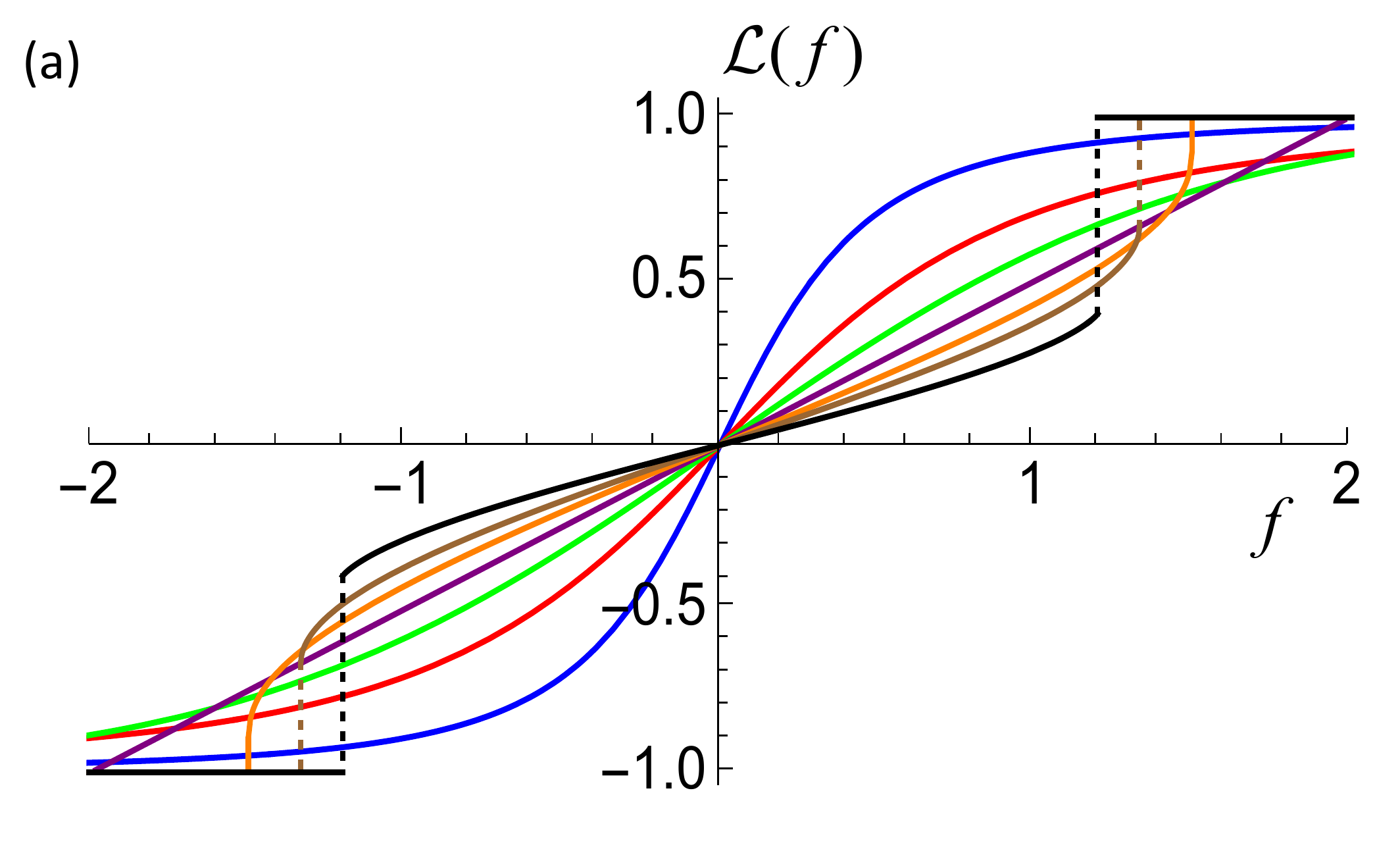}
\includegraphics[width=0.6\linewidth]{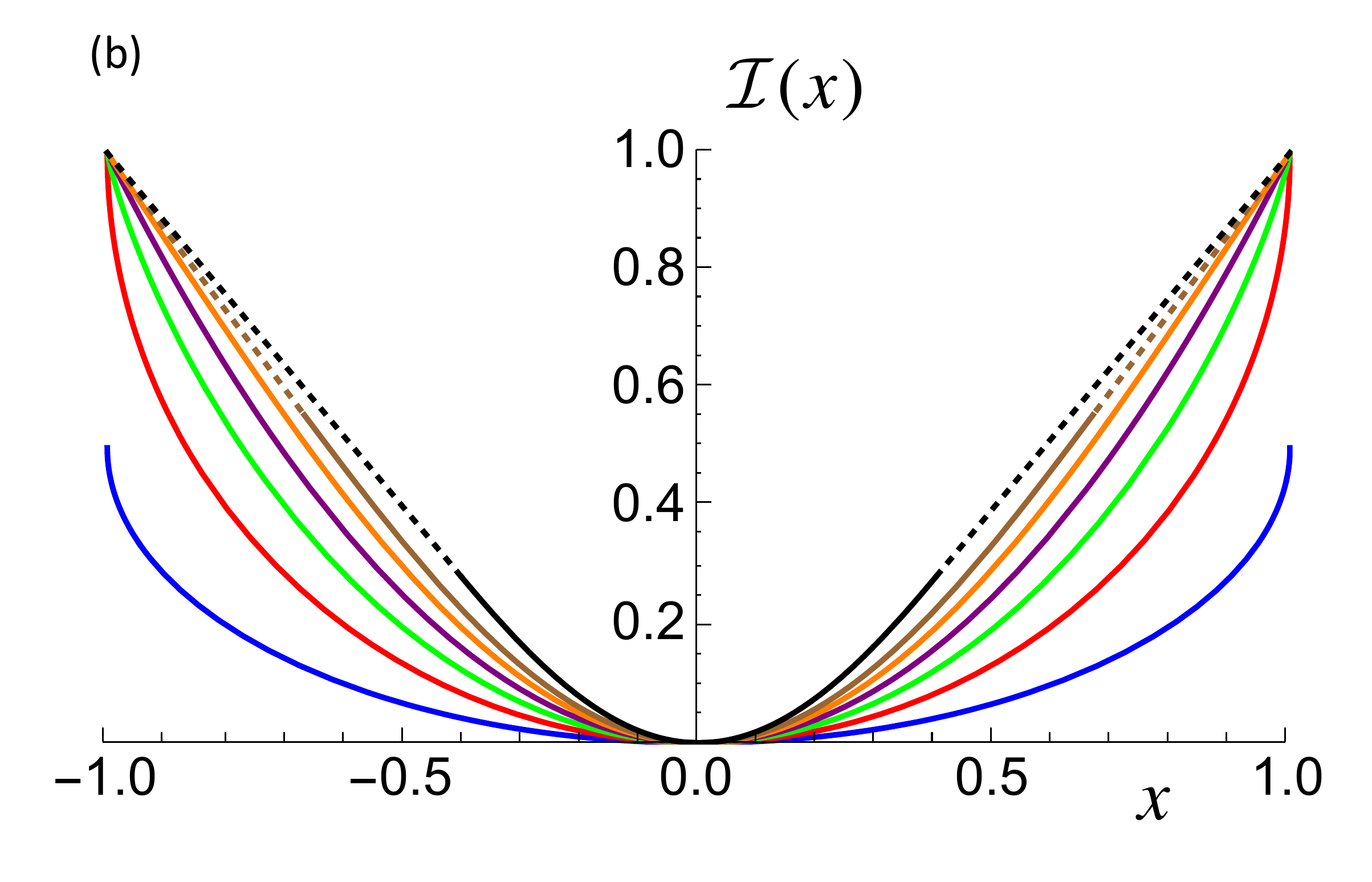}
\caption{a) Langevin function $\mathcal{L}(f)$ for $d=1,2,3,4,5,6,8$ (from decreasing slope at the origin) and b) large deviation function $\mathcal{I}(x)$ (same color code) for the run-and-tumble walk with $\ell_0=\tau_0=v_0=1$.}
\label{fig:OC}
\end{figure}

A further inspection of these results is called for, see also Fig.~\ref{fig:OC}. The LDF's for $d\le 5$ are smooth convex functions linking the point $\mathcal{I}(0)=0$ to $\mathcal{I}(\pm 1)=1$ ($\mathcal{I}(\pm 1)=1/2$ for $d=1$). The first surprise arises with $d=4$: the LDF is purely quadratic up to full extension $|x|=1$. This result was anticipated in \cite{karel} from a general criterion for such ``perfect harmonicity". Turning to $d=5$, we note that the LDF is still convex, but with inflection points at $|x|=1$. 
This signals the departure from convexity, which is apparent when turning to $d=6$: the inflection points are now at $|x|=x_\text{c}=2/3$; the LDF is no longer convex for $|x|>x_\text{c}$ and takes on the wrong limiting value $\mathcal{I}(\pm 1)=15/16$ instead of $\mathcal{I}(\pm 1)=1$. Furthermore, starting with $d=4$, the value of the generalized Langevin function ${\mathcal L}_d(f)$ seems to be greater than $1$ for large enough $f>f_\text{c}$, which does not make sense as $x={\mathcal L}_d(f)$ is a length restricted to the interval $x\in [-1,1]$. 

The reason for these inconsistencies is the same that was apparent in the off-lattice random walks in discrete time, namely that Eq.(\ref{eq:Cf1}) might not be able to provide the function ${\mathcal C}_d(f)$ for all values of $f$. The analysis is very similar to the one carried out in Subsection \ref{subsec:non-null}. The existence of integral (\ref{eq:Cf1}) requires $|1+{\mathcal C}_d(f)|>|f|$. There is a limiting value $f_\text{c}$, defined as $1+{\mathcal C}_d(f_\text{c})=f_\text{c}$ (we focus now only of $f>0$), such that the integral does not exist and ${\mathcal C}_d(f)$ can not be found using Eq.(\ref{eq:Cf1}). In this case it is possible to give an explicit value of $f_\text{c}$ using Eq.(\ref{eq:Fdz}) with $f=f_\text{c}$ that leads to
\begin{equation}
f_\text{c}=F\left(\frac12,1;\frac{d}{2};1\right)=\begin{cases}\infty, &d\le 3,\\ \dfrac{d-2}{d-3}, &d\ge 3.\end{cases}
\end{equation}

The solution is found, again, by considering explicitly the contribution of the $\theta=0$ direction which is neglected in Eq.(\ref{eq:Cf1}) for $d>3$ due to the weighting factor $g_d(\theta)$, i.e. by replacing, for $f>0$, Eq.(\ref{eq:Cf1}) with
\begin{equation}\label{eq:Cf2}
1=\frac{1}{M}\frac{1}{1+{\mathcal C}_d(f)-f}+\int d\theta\,\dfrac{g_d(\theta)}{1+{\mathcal C}_d(f)-f\cos(\theta)},
\end{equation}
which now admits a solution ${\mathcal C}_d(f)$ for all values of $f$. When $M\to\infty$, the solution for $f>f_\text{c}$ tends to ${\mathcal C}_d(f)=-1+f$ and, consequently, ${\mathcal L}_d(f)=1$.

The value of ${\mathcal L}_d(f_\text{c})$ can be obtained explicitly setting $z=1$ in Eq.(\ref{eq:param6}):
\begin{equation}
{\cal L}_d(f_\text{c})=\dfrac{2 F\left(\frac32,2;1+\frac{d}{2};1\right)}{dF\left(\frac12,1;\frac{d}{2};1\right)+2 F\left(\frac32,2;1+\frac{d}{2};1\right)}=\frac{2}{d-3}, d\ge 3.
\end{equation}

Hence, the situation is analogous to what happened in the off-lattice model with persistence. For $d\le 3$ the Langevin function ${\mathcal L}_d(f)$ tends to $\pm 1$ as $f\to\pm\infty$. For $3<d\le 5$, the value ${\mathcal L}_d(f)=\pm1$ is reached at a finite value $\pm f_\text{c}=\pm\dfrac{d-2}{d-3}$. For $d>5$, $|{\mathcal L}_d(\pm f_\text{c})|=\dfrac{2}{d-3}<1$, and there is a discontinuity as $|{\mathcal L}_d(f)|=1$ for $|f|>f_\text{c}$.

Concerning the LDF ${\mathcal I}_d(x)$, it turns out  that it becomes non-convex for $d>5$ and $x>x_\text{c}={\mathcal L}_d(f_\text{c})=\frac{2}{d-3}$, yielding an incorrect value ${\mathcal I}_d(x=\pm 1)<1$. Again, the solution comes by means of the Maxwell construction, connecting the $(x_\text{c},{\mathcal I}_d(x_\text{c}))$ and $(1,1)$ points by the straight line
\begin{equation}
{\mathcal I}_d(x)=1+\frac{d-2}{d-3}(|x|-1), \text{ for }~ d>5 \text{ and } x_\text{c}<|x|<1,
\end{equation}
that results from ${\mathcal I}_d(x_\text{c})=x_\text{c}f_\text{c}-{\mathcal C}_d(f_\text{c})=x_\text{c}f_\text{c}-(f_\text{c}-1)=(x_\text{c}-1)f_\text{c}+1$. This is the procedure that has been followed when plotting the LDF for $d=6,8$ in Fig.3(b).

For $|f| \geq f_\text{c}$, the linear segments in the LDF imply that the Langevin function becomes a constant, i.e., no increase in force is needed to achieve further elongation beyond the critical elongation $x_\text{c}$, see vertical dashed lines in Fig.~\ref{fig:OC}(a). The constancy of the force is a trademark of a first order phase transition, akin to constancy of the pressure during the gas-liquid transition. Concerning the nature of the random walks, fully straight flight segments of macroscopic length, i.e., proportional to $L$, will constitute a finite fraction of the realizations for end-to-end distances exceeding $|x|=x_\text{c}$ and our calculations allow us to calculate exactly the macroscopic weight of these segments.  Note that the values of $x_\text{c}$ are close to those observed in the persistent off-lattice random walk, while the values for $ f_\text{c}$ differ by a factor close to $\ln 2$. The phase transitions in the run-and tumble-limit and the persistent off-lattice cases are qualitatively but also to some extent quantitatively similar.

\section{Conclusions}
In summary, we encounter a number of surprising features that are absent in the walks without persistence. Firstly, on-lattice random walks with persistence, a pair of new inflection points (wiggles) appear in the Langevin function for $d\geq 3$, implying an initial phase of softening followed by the usual stiffening beyond a critical value of the force amplitude. In the limit of infinite dimension, the two wiggles turn into a pair of discontinuous transitions, with no extension below, and full extension above a critical value of the applied force. For off-lattice random walks with persistence, the large deviation function undergoes a first order phase transition in dimension $d>5$. This transition is of the "condensation type" in the sense that the occupancy of a single state (namely the persistent direction) becomes macroscopic beyond a certain extension. In other words, the condensed phase corresponds to a macroscopic fraction of the random walk oriented along the end-to-end distance. This is obviously reminiscent of a transition to crystallization. In the corresponding force-versus-extension relation, the extension becomes independent of the force beyond a critical value. The transition is anticipated in $d>3$, where full extension if reached at a finite value of the applied stretching force.

We make some final comments. First, we have chosen to discuss persistency as a purely entropic phenomenon. The resulting temperature independent phase transition arises, as in hard core liquids \cite{frenkel}, from a competition between two forms of entropy, one associated to persistency and the other to the space angle. One could attribute an energetic origin to persistency, which would result in a temperature dependent phase transition. Second, we expect that the critical dimension of the reported phase transition can be reduced by complementing the space angle entropic cost with an energetic cost induced by attractive forces. Thirdly, it was recently shown that dynamic phase-transitions also exist in a different model for persistent random walks \cite{rwpt}. It would be interesting to study the relation with our result. Fourthly, our random walk model can be mapped on a one-dimensional spin chain, where the direction of the spins corresponds to the direction of the walker. This mapping is particularly interesting, as the number of one-dimensional models with local interactions exhibiting phase-transitions is very limited \cite{pt1,pt2}. Finally, the observed phase transition, with the macroscopic appearance of fully stretched segments, is very reminiscent of a stretched-induced crystallization \cite{k,ct}, but further research is needed to clarify the exact correspondence.

\appendix

\section{Some analytical expressions}\label{analytical}
It turns out that for $d$ odd the integral Eq.(\ref{Fd:eq2}) can be expressed in terms of the polylogarithm\footnote{{\tt https://en.wikipedia.org/wiki/Polylogarithm}} $\Li_k(z)$ via the functions:
\begin{eqnarray}
G_k(\xi,f)&=&\dfrac{\Li_k(e^{-f}/\xi)+(-1)^k\Li_k(e^f/\xi)}{f^k},\\
G_k^-(\xi,f)&=&\dfrac{\Li_k(e^{-f}/\xi)-(-1)^k\Li_k(e^f/\xi)}{f^k},
\end{eqnarray}
as
\begin{eqnarray}
F_d(\xi,f)&=&\dfrac{\Gamma\left(\frac{d}{2}\right)}{\sqrt{\pi}\Gamma\left(\frac{d-1}{2}\right)}(-1)^{b+1}\sum_{k=0}^b2^{b-k}{b \choose k}(b+k)! G_{b+k+1}(\xi,f),\nonumber \\
b&=&\frac{d-3}{2} \text{ integer.}
\end{eqnarray}
Specific cases:
\begin{eqnarray}
F_3(\xi,f)&=&-\frac12G_1(\xi,f),\\
F_5(\xi,f)&=&\frac{3}{2}\left[G_2(\xi,f)+G_3(\xi,f)\right],\\
F_7(\xi,f)&=&-\frac{15}{2}(G_3(\xi,f)+3G_4(\xi,f)+3G_5(\xi,f)),\\
F_9(\xi,f)&=&\frac{105}{2}(G_4(\xi,f)+6G_5(\xi,f)+15G_6(\xi,f)+15G_7(\xi,f)).
\end{eqnarray}

The function ${\mathcal L}_d(f)=\dfrac{\xi'_d(f)}{\xi_d(f)}$ is obtained taking the derivative of $F_d(\xi_d(f),f)=\dfrac{{\mathcal P}}{1-{\mathcal P}}$ with respect to $f$:
\begin{eqnarray}
\frac{\partial F_d(\xi_d,f)}{\partial \xi_d}\frac{d\xi_d(f)}{df}+\frac{\partial F_d(\xi_d,f)}{\partial f}=0\Rightarrow {\mathcal L}_d(f)=\left.-\frac{1}{\xi}\dfrac{\left(\frac{\partial F_d(\xi,f)}{\partial f}\right)}{\left(\frac{\partial F_d(\xi,f)}{\partial \xi}\right)}\right|_{\xi\to \xi_d(f)}
\end{eqnarray}
As the derivative of the polylogarithm function can be expressed in terms of polylogarithm functions, it is possible to derive explicit expressions for $d$ odd as:
\begin{eqnarray}
{\mathcal L}_5(f)&=&\left.-\frac{G_1(\xi,f)+3G_2(\xi,f)+3G_3(\xi,f)}{G_1^-(\xi,f)+G_2^-(\xi,f)}\right|_{\xi\to \xi_d(f)},\\ \nonumber \\
{\mathcal L}_7(f)&=&\left.-\frac{G_2(\xi,f)+6G_2(\xi,f)+15G_4(\xi,f)+15G_5(\xi,f)}{G_2^-(\xi,f)+3G_3^-(\xi,f)+3G_4^-(\xi,f)}\right|_{\xi\to \xi_d(f)},\\ \nonumber \\
{\mathcal L}_9(f)&=&\left.-\frac{G_3(\xi,f)+10G_4(\xi,f)+45G_5(\xi,f)+105G_6(\xi,f)+105G_7(\xi,f)}{G_3^-(\xi,f)+6G_4^-(\xi,f)+15G_5^-(\xi,f)+15G_6^-(\xi,f)}\right|_{\xi\to \xi_d(f)}.\\ \nonumber
\end{eqnarray}

\section{Alternative derivations of the large deviation function for the run-and-tumble model}\label{alternative}

\subsection{Direct approach}
The probability (density) of having a coordinate $X$ at time $t$ satisfies:
\begin{equation}\label{PXt}
P(X,t)=\int_0^\infty d\tau\, \int_0^\pi d\theta\, e^{-\tau}g_d(\theta)P(X-\tau\cos\theta,t-\tau),
\end{equation}
and this is expected to follow a large deviation form $P(X,t)\sim e^{-t {\mathcal I}_d(X/t)}$.

The generating function $G(f,t)=\left\langle e^{fX}\right\rangle=\int dX P(X,t) e^{fX}$ yields the Legendre-Fenchel transform ${\mathcal C}_d(f)=\lim_{t\to\infty}\frac{1}{t}\ln G(f,t)$ of the large deviation function ${\mathcal I}_d(x)$. Inserting Eq.(\ref{PXt}) in the definition of $G(f,t)$ we arrive at:
\begin{equation}
G(f,t)=\int d\tau\, d\theta\, g_d(\theta)e^{-\tau(1-f\cos\theta)}G(f,t-\tau).
\end{equation}
We insert the ansatz $G(f,t)=e^{t{\mathcal C}_d(f)}$ and perform the integral over $\tau$, leading again to Eq.(\ref{eq:Cf1}).

\subsection{Master equation approach}
Let us start from a finite set $\theta_1,\dots,\theta_M$ of possible orientations to choose from and let $P(X,\theta_i,t)$ be the probability that the walker is at X-coordinate $X$ and has reached it from the $\theta_i$ direction. This probaiblity satisfies a master equation than can be derived from:
\begin{equation}
P(X,\theta_i,t+dt)=P(X-dt\cos\theta_i,\theta_i,t)(1-dt)+\frac{dt}{M}\sum_{j=1}^MP(X-dt\cos\theta_i,\theta_j,t).
\end{equation}
Expanding at first order in $dt$ and taking the limit $dt\to 0$ one gets:
\begin{equation}
\frac{\partial P(X,\theta_i,t)}{\partial t}=-P(X,\theta_i,t)-\cos\theta_i\frac{\partial P(X,\theta_i,t)}{\partial X}+\frac{1}{M}\sum_j P(X,\theta_j,t).
\end{equation}
For the set of generating functions $G(f,\theta_i,t)=\int dX e^{fX}P(X,\theta_i,t)$ we obtain after an integration per parts:
\begin{equation}
\frac{\partial G(f,\theta_i,t)}{\partial t}=-G(f,\theta_i,t)+f\cos\theta_iG(f,\theta_i,t)+\frac{1}{M}\sum_jG(f,\theta_j,t).
\end{equation}
A set of $M$ coupled linear differential equations for which we try the ansatz $G(f,\theta_i,t)=\Psi_i(f)e^{t{\mathcal C}_d(f)}$, or:
\begin{equation}
{\mathcal C}_d\Psi_i=(-1+f\cos\theta_i)\Psi_i+\frac{1}{M}\sum_j\Psi_j.
\end{equation}
The algebra is now similar to the one used in the study of random walks on a lattice with persistence and leads to 
\begin{equation}
1=\frac{1}{M}\sum_{j=1}^M\frac{1}{1+{\mathcal C}_d-f\cos\theta_j},
\end{equation}
which in the limit $M\to\infty$ recovers Eq.(\ref{eq:Cf1}).

\subsection{Using contraction theorem}

Our starting point here is the explicit, exact result for the large deviation function of the empirical distribution for a Markov process obeying detailed balance. For large times $t$, the empirical distribution converges to the genuine probability $p(\Omega)$. This convergence is described by a LDF $ J[q(\Omega)]$. Note that $J$ is a functional whenever the state space $\Omega$ is continuous since $q(\Omega)$ is then a function. Our basic starting point is that the explicit form of this LDF is known for a continuous time Markov process obeying detailed balance \cite{Hollander:2008,Donsker-Varadhan:1975}, namely:
\begin{eqnarray}
 J[q(\Omega)]&=&\int d\Omega_1\int d\Omega_2\frac{1}{2}\left(\sqrt{W({\Omega_2,\Omega_1})q({\Omega_1})}-\sqrt{W({\Omega_1,\Omega_2})q({\Omega_2})}\right)^2\nonumber\\&=&\int d\Omega_1\int d\Omega_2W({\Omega_2,\Omega_1})q({\Omega_1})\nonumber\\&&-\int d\Omega_1\int d\Omega_2\sqrt{W({\Omega_1,\Omega_2})W({\Omega_2,\Omega_1})q({\Omega_1})q({\Omega_2})},\nonumber\\\label{LDFJ}
\end{eqnarray}
where $W({\Omega_1,\Omega_2})$ is the transition function, and the summation (or integration) runs over the full space of $\Omega_1$ and $\Omega_2$. In our system, we have $W({\Omega_1,\Omega_2})=1/\int d\Omega$, i.e.~a uniform transition rate. This leads to
\begin{equation}
 J[q(\Omega)]=1-\frac{\left(\int d\Omega \sqrt{q(\Omega)}\right)^2}{\int d\Omega}
\end{equation}

For its application to our random walk problem, we assume that the speed of progression along a direction $x$ is given by a function $f({\Omega})=\cos(\theta)$, where $\theta$ is the angle between the preferred direction and the direction of $\Omega$.
The LDF $\mathcal{I}(x)$ for $x$ is then obtained by a ``contraction". More specifically, it is given by:
\begin{equation}\label{LDFv}
\mathcal{I}(x)=J[\bar{q}(\Omega)],
\end{equation} 
where $\bar{q}$ is the most likely empirical distribution, i.e. the one that minimized the LDF $J[{q}(\Omega)]$, while satisfying the constraint that it realizes the correct value of $x$, while of course obeying normalization:
\begin{eqnarray}
 \int d\Omega \bar{q}(\Omega)&=&1,\label{con2}\\
 \int d\Omega \bar{q}(\Omega)\cos(\theta)&=&x.\label{con1}
\end{eqnarray}
Note that we have, for simplicity of notation, omitted the dependence of $\bar{q}(\Omega)$ on $x$, $\bar{q}(\Omega)=\bar{q}_x(\Omega)$ . This dependence is in fact crucial as it reveals the empirical distribution that will be observed with exponentially overwhelming probability in the realizations that correspond to an observed $x$.

The above minimization can be performed using Lagrange multipliers. The function $\bar{q}(\Omega)$ that minimizes the expression
\begin{equation}
J[q(\Omega)]+(1+c_1)\left(\int d\Omega q(\Omega)-1\right)-c_2\left(\int d\Omega q(\Omega)f({\Omega})-x\right).
\end{equation}
(the arbitrary multipliers $1+c_1$ and $-c_2$ simplify some later formulas)
is found from basic variational calculus to obey the following integral equation:
\begin{equation}
\bar{q}(\Omega)=\left(\frac{\int d\Omega' \sqrt{\bar{q}({\Omega'})}}{\left(1+c_1+c_2 \cos(\theta)\right)\int d\Omega'}\right)^2.\label{qeq}
\end{equation}
 This equation has to be solved together with the constraints Eqs.~(\ref{con1})-(\ref{con2}). The LDF $\mathcal{I}(x)$ for the sample speed follows by plugging this result for $\bar{q}(\Omega)$ into Eq.~(\ref{LDFv}).
 
One can rewrite Eq.~(\ref{qeq}) as
\begin{equation}
\bar{q}(\Omega)=\frac{r}{\left(1+c_1-c_2 \cos(\theta)\right)^2\int d\Omega},\label{qres}
\end{equation}
with
\begin{equation}
 \sqrt{r}=\frac{\int d\Omega \sqrt{\bar{q}(\Omega)}}{\sqrt{\int d\Omega}}.\label{rdef}
\end{equation}
With this definition one can easily check that
\begin{equation}\label{idef}
 {\mathcal I}_d=1-r
\end{equation}
Furthermore, plugging Eq.~(\ref{qres}) into Eqs.~(\ref{rdef}),(\ref{con2}) and (\ref{con1}) leads respectively to
\begin{eqnarray}
S_d(c_1 ,c_2)&=&1,\label{eq1}\\
-r\dfrac{\partial S_d(c_1 ,c_2)}{\partial c_1 }&=&1,\label{eq2}\\
r\dfrac{\partial S_d(c_1 ,c_2)}{\partial c_2}&=&x,\label{eq3}
\end{eqnarray}
where 
\begin{equation}
S_d(c_1 ,c_2)=\int_0^\pi d\theta \dfrac{g_d(\theta)}{1+c_1 -c_2\cos(\theta)}.
\end{equation}

The similarity of this relation with Eq.(\ref{eq:Cf1}) indicates that $c_1(c_2)$ follows the same functional relation as ${\mathcal C}_d(f)$. Furthermore, using Eqs.(\ref{eq2},\ref{eq3}) we obtain $\dfrac{dc_1}{dc_2}=x$, confirming that $c_1$ can be identified with ${\mathcal C}_d$ and $c_2$ with $f$.

\section*{Acknowledgements}

{KP is a postdoctoral fellow of the Research Foundation-Flanders (FWO). RT acknowledges financial support from Agencia Estatal de Investigaci\'on (AEI, Spain) and Fondo Europeo de Desarrollo Regional under Grant No. FIS2015-63628-C2-2-R (AEI/FEDER,UE) and the Spanish State Research Agency, through the Mar{\'\i}a de Maeztu Program for Units of Excellence in R\&D (MDM-2017-0711).}


\begin{thebibliography}{99}
\bibitem{tp}
The Gibbs free energy is given by the Euler relation
$\mathcal{G}=\mathcal{U} - T\mathcal{S} - FX$. In the case considered here, the internal energy $\mathcal{U}$ is identically zero.
\end{thebibliography}
\end{document}